\documentclass[english]{elsart}
\usepackage{babel}
\input{epsf}
\begin{document}

\newcommand{\cerenkov}{\v{C}erenkov }
\newcommand{\pythia}{\textsc{Py\-thia/Jet\-set} }
\newcommand{\pythi}{\textsc{Py\-thia$5.7$/Jet\-set$7.4$} }
\newcommand{\pyt}{\textsc{Py\-thia} }
\newcommand{\jet}{\textsc{Jet\-set} }
\def\ss{\scriptscriptstyle}
\def\pbar{\overline{p}}
\def\Lm{\Lambda^{\circ}}
\def\Lbar{\overline{\Lambda}^{\circ}}
\def\Xim{\Xi^-}
\def\Xibar{\overline{\Xi}^+}
\def\Om{\Omega^-}
\def\Ombar{\overline{\Omega}^+}
\def\Lc{\Lambda_{\ss c}^+}
\def\Lcbar{{\Lambda}_{\ss c}^{\ss -}}
\def\Lcc{\Lambda_{\ss c}}

\def\Np{N^{\prime}}
\def\lcsub{_{\scriptscriptstyle{\Lc}}}
\def\bbarsub{_{\scriptscriptstyle \overline{B}}}
\def\xf{x_{\scriptscriptstyle F}}
\def\pt{p_{\scriptscriptstyle T}^2}

\def\etal{{\it et al.}}
\def\PL#1{Phys. Lett. {\bf#1}}
\def\PR#1{Phys. Rev. {\bf#1}}
\def\PRL#1{Phys. Rev. Lett. {\bf#1}}
\def\NP#1{Nucl. Phys. {\bf#1}}
\def\NIM#1{Nucl. Instr. and Methods {\bf#1}}

\hyphenation{pa-ra-me-tri-za-tion}

\begin{frontmatter}
                      
\title{Asymmetries in the Production of $\Lc$ and $\Lcbar$ Baryons in 500 GeV/c
 $\pi^-$ Nucleon Interactions}

%
\collab {Fermilab E791 Collaboration}

\author[umis]{E.~M.~Aitala},
\author[cbpf]{S.~Amato},
\author[cbpf]{J.~C.~Anjos},
\author[fnal]{J.~A.~Appel},
\author[taun]{D.~Ashery},
\author[fnal]{S.~Banerjee},
\author[cbpf]{I.~Bediaga},
\author[umas]{G.~Blaylock},
\author[stev]{S.~B.~Bracker},
\author[stan]{P.~R.~Burchat},
\author[ilit]{R.~A.~Burnstein},
\author[fnal]{T.~Carter},
\author[cbpf]{H.~S.~Carvalho},
\author[usca]{N.~K.~Copty},
\author[umis]{L.~M.~Cremaldi},
\author[yale]{C.~Darling},
\author[fnal]{K.~Denisenko},
\author[ucin]{S.~Devmal},
\author[pueb]{A.~Fernandez},
\author[usca]{G.~F.~Fox},
\author[ucsc]{P.~Gagnon},
\author[cbpf]{C.~Gobel},
\author[umis]{K.~Gounder},
\author[fnal]{A.~M.~Halling},
\author[cine]{G.~Herrera},
\author[taun]{G.~Hurvits},
\author[fnal]{C.~James},
\author[ilit]{P.~A.~Kasper},
\author[fnal]{S.~Kwan},
\author[usca]{D.~C.~Langs},
\author[ucsc]{J.~Leslie},
\author[fnal]{B.~Lundberg},
\author[cbpf]{J.~Magnin},
\author[taun]{S.~MayTal-Beck},
\author[ucin]{B.~Meadows},
\author[cbpf]{J.~R.~T.~de~Mello~Neto},
\author[ksun]{D.~Mihalcea},
\author[tuft]{R.~H.~Milburn},
\author[cbpf]{J.~M.~de~Miranda},
\author[tuft]{A.~Napier},
\author[ksun]{A.~Nguyen},
\author[ucin,pueb]{A.~B.~d'Oliveira},
\author[ucsc]{K.~O'Shaughnessy},
\author[ilit]{K.~C.~Peng},
\author[ucin]{L.~P.~Perera},
\author[usca]{M.~V.~Purohit},
\author[umis]{B.~Quinn},
\author[uwsc]{S.~Radeztsky},
\author[umis]{A.~Rafatian},
\author[ksun]{N.~W.~Reay},
\author[umis]{J.~J.~Reidy},
\author[cbpf]{A.~C.~dos Reis},
\author[ilit]{H.~A.~Rubin},
\author[umis]{D.~A.~Sanders},
\author[ucin]{A.~K.~S.~Santha},
\author[cbpf]{A.~F.~S.~Santoro},
\author[ucin]{A.~J.~Schwartz},
\author[cine,uwsc]{M.~Sheaff},
\author[ksun]{R.~A.~Sidwell},
\author[cbpf]{F.~R.~A.~Sim\~ao},
\author[yale]{A.~J.~Slaughter},
\author[ucin]{M.~D.~Sokoloff},
\author[cbpf]{J.~Solano},
\author[ksun]{N.~R.~Stanton},
\author[fnal]{R.~J.~Stefanski},
\author[uwsc]{K.~Stenson},  
\author[umis]{D.~J.~Summers},
\author[yale]{S.~Takach},
\author[fnal]{K.~Thorne},
\author[ksun]{A.~K.~Tripathi},
\author[uwsc]{S.~Watanabe},
\author[taun]{R.~Weiss-Babai},
\author[prin]{J.~Wiener},
\author[ksun]{N.~Witchey},
\author[yale]{E.~Wolin},
\author[ksun]{S.~M.~Yang},
\author[umis]{D.~Yi},
\author[ksun]{S.~Yoshida},
\author[stan]{R.~Zaliznyak}, and
\author[ksun]{C.~Zhang}

\address[cbpf]{Centro Brasileiro de Pesquisas F{\'\i}sicas, Rio de Janeiro, Brazil}
\address[ucsc]{University of California, Santa Cruz, California 95064, USA}
\address[ucin]{University of Cincinnati, Cincinnati, Ohio 45221, USA}
\address[cine]{CINVESTAV, 07000 Mexico City, DF Mexico}
\address[fnal]{Fermilab, Batavia, Illinois 60510, USA}
\address[ilit]{Illinois Institute of Technology, Chicago, Illinois 60616, USA}
\address[ksun]{Kansas State University, Manhattan, Kansas 66506, USA}
\address[umas]{University of Massachusetts, Amherst, Massachusetts 01003, USA}
\address[umis]{University of Mississippi--Oxford, University, Mississippi 38677, USA}
\address[prin]{Princeton University, Princeton, New Jersey 08544, USA}
\address[pueb]{Universidad Autonoma de Puebla, Mexico}
\address[usca]{University of South Carolina, Columbia, South Carolina 29208, USA}
\address[stan]{Stanford University, Stanford, California 94305, USA}
\address[taun]{Tel Aviv University, Tel Aviv 69978, Israel}
\address[stev]{Box 1290, Enderby, British Columbia V0E 1V0, Canada}
\address[tuft]{Tufts University, Medford, Massachusetts 02155, USA}
\address[uwsc]{University of Wisconsin, Madison, Wisconsin 53706, USA}
\address[yale]{Yale University, New Haven, Connecticut 06511, USA}

\begin{abstract}

We present a measurement of asymmetries in the production of 
$\Lc$ and $\Lcbar$ baryons in 500 GeV/$c$
$\pi^-$--nucleon interactions from the E791 experiment at Fermilab. 
The asymmetries were measured as functions of Feynman $x$ ($\xf$) and 
transverse momentum squared ($\pt$) using a sample of $1\,819 \pm 62$ 
$\Lcc$'s observed in the decay channel $\Lc\rightarrow pK^-\pi^+$. 
We observe more $\Lc$ than $\Lcbar$ baryons, with an asymmetry of 
$(12.7\pm3.4\pm1.3)\;\%$ independent of $\xf$ and $\pt$ in our kinematical 
range ($-0.1\leq \xf \leq 0.6$ and $0.0\leq\pt\leq 8.0 ~({\rm GeV/c})^2$).
This $\Lcc$ asymmetry measurement is the first with data in 
both the positive and negative $\xf$ regions.

\end{abstract}
\end{frontmatter}

Several studies \cite{mesons} have reported an 
enhancement in the production rate 
of charm particles having a valence quark or diquark in common with the 
incident particles, relative to charge-conjugate particles which have 
fewer or no common valence quarks. This effect, known as the 
leading-particle effect, cannot be accounted for by 
next-to-leading-order perturbative QCD (pQCD) calculations followed 
by Peterson fragmentation of the produced $c$ and $\bar{c}$ quarks. 
Thus, the effect is attributed to additional hadronization 
effects or non-pQCD contributions to charm production.

In this work we report on the $\Lc - \Lcbar$ production 
asymmetry measured using a sample containing $1\,819\pm 62$ 
fully-reconstructed $\Lc \rightarrow pK^-\pi^+$ 
and charge conjugate (c.c.) decays from the Fermilab 
E791 experiment. The data are from 500 GeV/$c$ $\pi^-$-nucleon 
interactions. This measurement represents a significant increase in 
statistics from previously published results, and it also explores for the 
first time the production asymmetry at negative values of $\xf$.
This feature of our measurement allows us to look both for 
diquark effects which should be present in the negative $\xf$ region (since  
the $\Lc$ has two valence quarks in common with the target particles whereas 
the $\Lcbar$ has none) 
and for leading particle effects in the positive $\xf$ 
region (in which both the particle and antiparticle share a single 
valence quark with the incident pions). The production asymmetries are 
measured as functions of $\xf$ and $\pt$ in the kinematic 
ranges $-0.1\leq\xf\leq 0.6$ and $0\leq\pt\leq 8~({\rm GeV/c})^2$.

Varied predictions about what asymmetries should be observed in the E791
data are made by three models which address leading particle effects:
the string fragmentation (SF) model \cite{lund}, the intrinsic charm (IC) model
\cite{ic}, and the two-component recombination (2CR) model \cite{rec}.

In the string fragmentation model implemented in the Lund \pythia 
package \cite{pyth}, the $c$ and $\bar{c}$ quarks produced by leading 
order QCD diagrams are connected
by color strings to valence quarks or diquarks in the beam or target
particles. Hadronization proceeds through fragmentation of the ends of the  
color string and recombination with valence quarks from the initial
particles.  The model predicts a negative asymmetry (more $\Lcbar$ than $\Lc$) for
$\xf > 0$, a positive asymmetry for $\xf < 0$, and a flat dependence on $\pt$.

For the intrinsic charm model, fluctuation of one or the other of the
incident particles to a Fock state containing $c\bar{c}$ quarks 
($\left|uudc\bar{c}\right>$ or $\left|uddc\bar{c}\right>$ for nucleons 
and $\left|\bar{u}dc\bar{c}\right>$ for $\pi^-$) favors
production of particles with valence quarks in common with the Fock state
via a coalescence (recombination) mechanism. Hence, this model predicts no
asymmetry in the $\xf > 0$ region and a positive asymmetry for $\xf < 0$,
increasing as one moves toward more negative $\xf$.

In the two-component recombination model, hadronization takes place
through fragmentation of perturbatively produced $c$ and $\bar{c}$ quarks, and 
the recombination of valence and $c$ ($\bar{c}$) quarks produced in the initial
collision. In the $\xf < 0$ region, this model predicts a positive
asymmetry growing strongly as $\xf$ becomes more negative, due to
target-diquark effects. Again, there is no asymmetry in the pion
fragmentation region ($\xf > 0$) since both $\Lc$ and $\Lcbar$ are singly-leading
particles.

The main difference between the IC and 2CR models is that the former
predicts very little asymmetry in the E791 negative $\xf$ range 
because the IC distribution
peaks at about $0.6$ of the initial particle momentum. Because of diquark
effects, the 2CR model can account for larger asymmetries in this region.

Neither the IC nor the 2CR model has been used to make a prediction
regarding the variation of asymmetry with respect to $\pt$.  Whatever
asymmetries do exist, however, are expected to decrease as $\pt$ increases,
since the density of beam and target valence quarks available for
recombination decreases with increasing $\pt$.

Experiment E791 recorded data from 500 ${\rm GeV/c}$ $\pi^-$ 
interactions in five thin foils (one platinum and four diamond) 
with center-to-center separations of about 15 mm. Each foil was 
approximately 0.4\% of 
a pion interaction length thick (0.5 mm for the platinum foil and 1.6 mm 
for the carbon foils). The E791 spectrometer \cite{e791} in the Tagged 
Photon Laboratory was a large-acceptance, two-magnet spectrometer 
with eight planes of multiwire proportional chambers (MWPC) and 
six planes of silicon 
microstrip detectors (SMD) placed upstream of the target 
for beam tracking. Downstream of the target 
were 17 planes of SMD's for track and vertex reconstruction,
35 drift chamber planes, two MWPC's, two multicell threshold
\cerenkov counters, electromagnetic and hadronic calorimeters, and a
muon detector. The experiment recorded $2 \times 10^{10}$ interactions using an 
open transverse-energy trigger and a fast data acquisition 
system \cite{viteum}.


A sample of candidate $\Lc\rightarrow pK^-\pi^+$ and 
c.c. decays was selected using similar criteria to those reported 
in Ref.~\cite{e791-lamc-sigc}. Since the mean decay length for $\Lcc$'s 
was several mm, in most cases they decayed in air gaps between the target 
foils and before entering the silicon 
vertex detectors. We selected all combinations of three tracks which had 
\cerenkov counter identification probabilities consistent with the 
$pK\pi$ hypothesis and which had a $pK\pi$ invariant mass between 
$2.15$ and $2.45$ GeV/$c\,^2$. 
We then required that the distance between the reconstructed candidate $\Lcc$ 
decay vertex and the production vertex be at least five times the 
rms uncertainty in that distance. 
To further reduce background, we required the $\Lambda_c$ to decay 
a distance greater than $5\sigma_L$ from the nearest 
material, where $\sigma_L$ is the uncertainty on the 
vertex $z$ position. The vertex position is also required to lie 
between one and four $\Lcc$ lifetimes from the primary (interaction) vertex. 
The $\Lcc$ momentum vector, reconstructed from its decay 
products, was required to pass within $50\,\mu m$ of the production 
vertex, and the reconstructed production and decay 
vertices were each required to have 
an acceptable $\chi ^2$. We also required that no more than one of the 
$\Lcc$-candidate decay tracks be consistent with coming from the 
primary interaction point. The sum of the $\pt$ of the secondary tracks with 
respect to the flight path of the reconstructed $\Lambda_c$ was required to be 
greater than $0.35~({\rm GeV/c})^2$ if $-0.1\leq \xf \leq 0.3$, and greater than 
$0.32~({\rm GeV/c})^2$ if $0.3< \xf \leq 0.6$. 

\begin{figure}[t]
\centerline{\epsfxsize=6.5in \epsffile{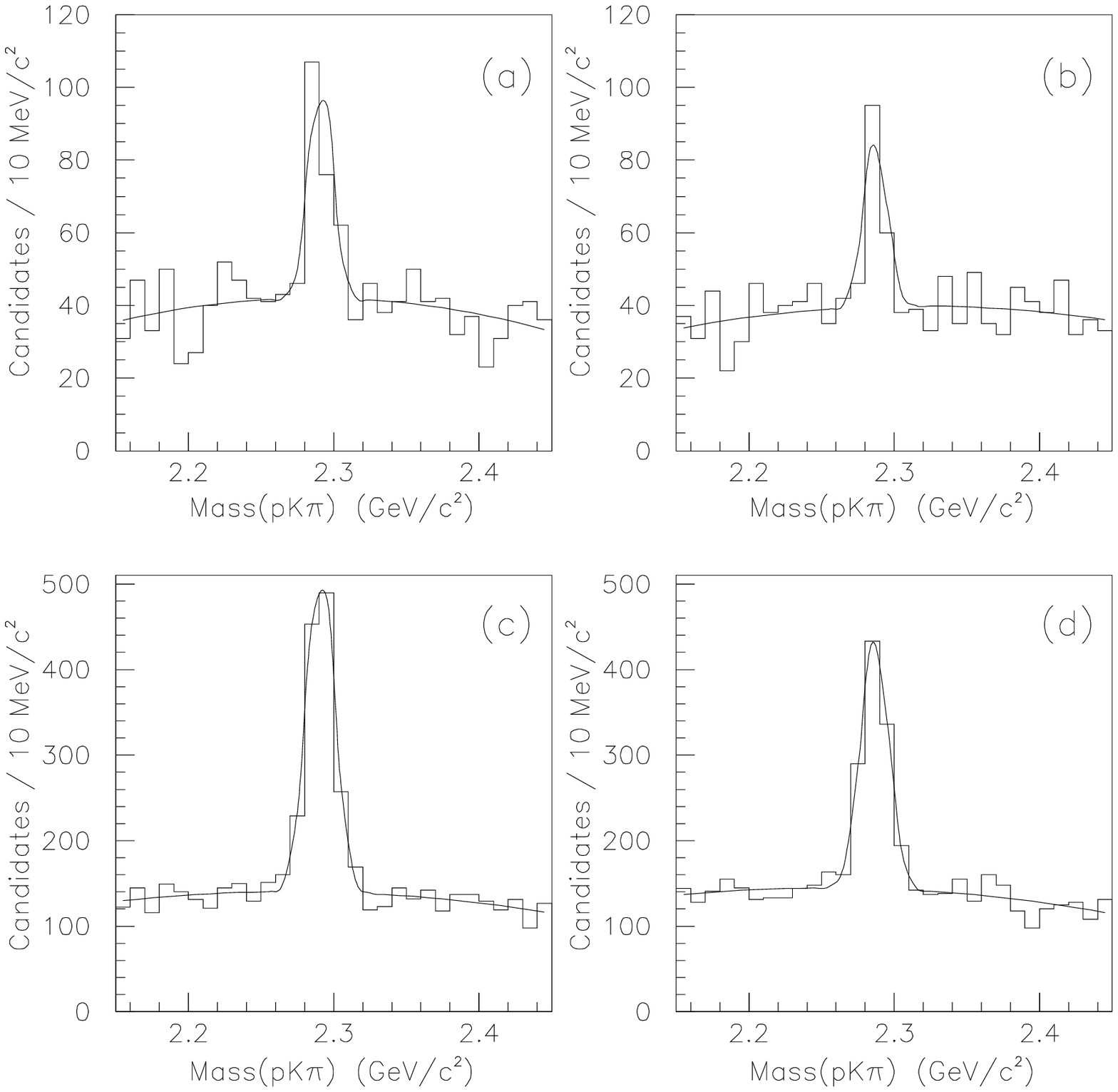}}
  \caption{$pK^-\pi^+$ and $\pbar K^+\pi^-$ invariant-mass distributions 
for $\xf<0$ and $\xf>0$. All distributions show a clear 
$\Lc$ and $\Lcbar$ signal. Fitting to a Gaussian signal and 
quadratic background yields $122 \pm 17$ $\Lc$ (a) and $92 \pm 15$ $\Lcbar$ (b) in 
the negative $\xf$ region and $903 \pm 42$ $\Lc$ (c) and $702 \pm 40$ $\Lcbar$ 
(d) in the positive $\xf$ region.}
  \label{fig1}
\end{figure}

Events were kept if the $\Lcc$ candidate had 
$-0.1\leq \xf\leq 0.6$ and $\pt\leq 8~({\rm GeV/c})^2$. 
The $pK\pi$ invariant-mass plots for the final $\Lc$ and $\Lcbar$ 
samples are shown in Figs.~\ref{fig1}(a) and (b) respectively for 
$\xf<0$, and in Figs.~\ref{fig1}(c) and (d) for $\xf>0$. 
Fits to a Gaussian signal and a quadratic background 
yield $1\,025\pm 45~\Lc\rightarrow pK^-\pi^+$ and 
$794\pm 42~\Lcbar\rightarrow\pbar K^+\pi^-$ in the entire $\xf$ and 
$\pt$ range. 


For each bin of $\xf$ or $\pt$ we define an asymmetry parameter $A$ as
\begin{equation}
A \equiv {N - \overline{N}/r \over N + \overline{N}/r}~;~~~
r = {\overline{\epsilon} \over \epsilon}
\end{equation}
where $N$ ($\overline{N}$) is the number of $\Lc$ ($\Lcbar$)
produced in the bin and the efficiency $\epsilon$ ($\overline{\epsilon}$) 
is the product of the geometrical acceptance and reconstruction efficiency  
for the bin.

\begin{figure}[t]
\centerline{\epsfxsize=6.5in \epsffile{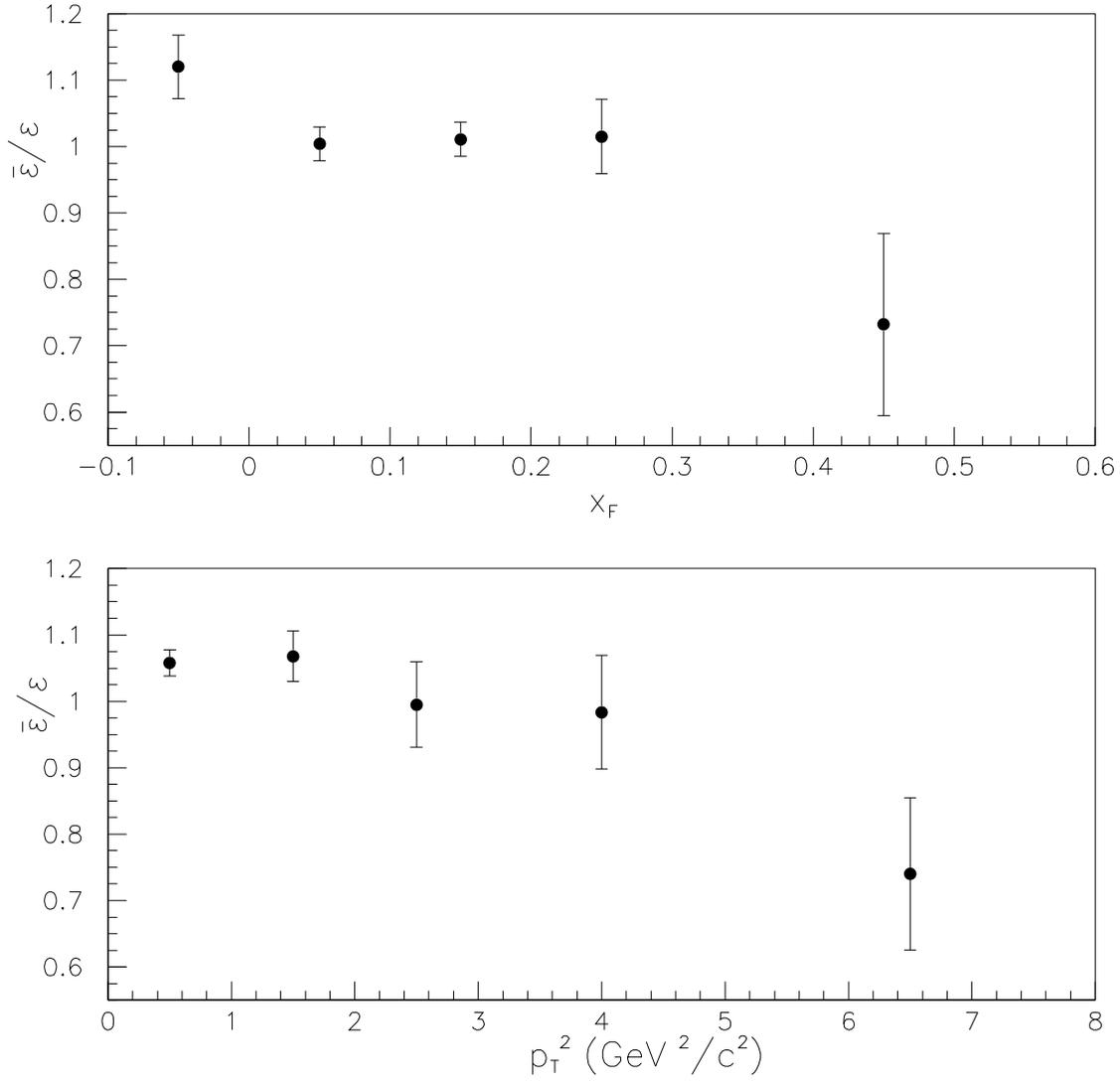}}
\caption{$\Lc$ and $\Lcbar$ efficiency ratios, $\bar{\epsilon}/
\epsilon$ as a function of $\xf$ (upper plot) and $\pt$ (lower plot). The efficiency 
ratios for $x_F$ ($\pt$) are integrated over the $\pt$ ($x_F$) range of 
the data set.}
\label{fig2}
\end{figure}

Values for $N$ ($\overline{N}$) were obtained from fits to $pK\pi$ 
invariant-mass distributions  for events within specific 
$\xf$ and $\pt$ ranges. In each case, well defined $\Lcc$ signals were evident. 
Efficiencies ($\epsilon$, $\bar{\epsilon}$) were calculated using a 
sample of $7\times 10^6$ Monte Carlo (MC) events produced with the \pythia event 
generator \cite {pyth}. These events were projected 
through a detailed simulation of the E791 detector and then 
reconstructed with the algorithms used for the data. The final 
reconstructed MC sample was approximately 
ten times the size of our data sample. In the simulation of the 
detector, special care was taken to accurately simulate the reduced 
chamber efficiency seen in the experiment in a 
small region of the tracking chambers nearest the beam. 
The behavior of the apparatus and details of the reconstruction
code changed during the data taking and data processing periods,
respectively. To account for these effects, we generated 
the final MC sample in subsets mirroring these behaviors and 
combined the samples using their known fractional contributions 
to the data set. 
Good agreement between MC and data samples in a variety of kinematic variables 
and resolutions was achieved. The efficiency ratios obtained 
are shown in Fig. \ref{fig2}.

\begin{figure}[t]
\centerline{\epsfxsize=6.5in \epsffile{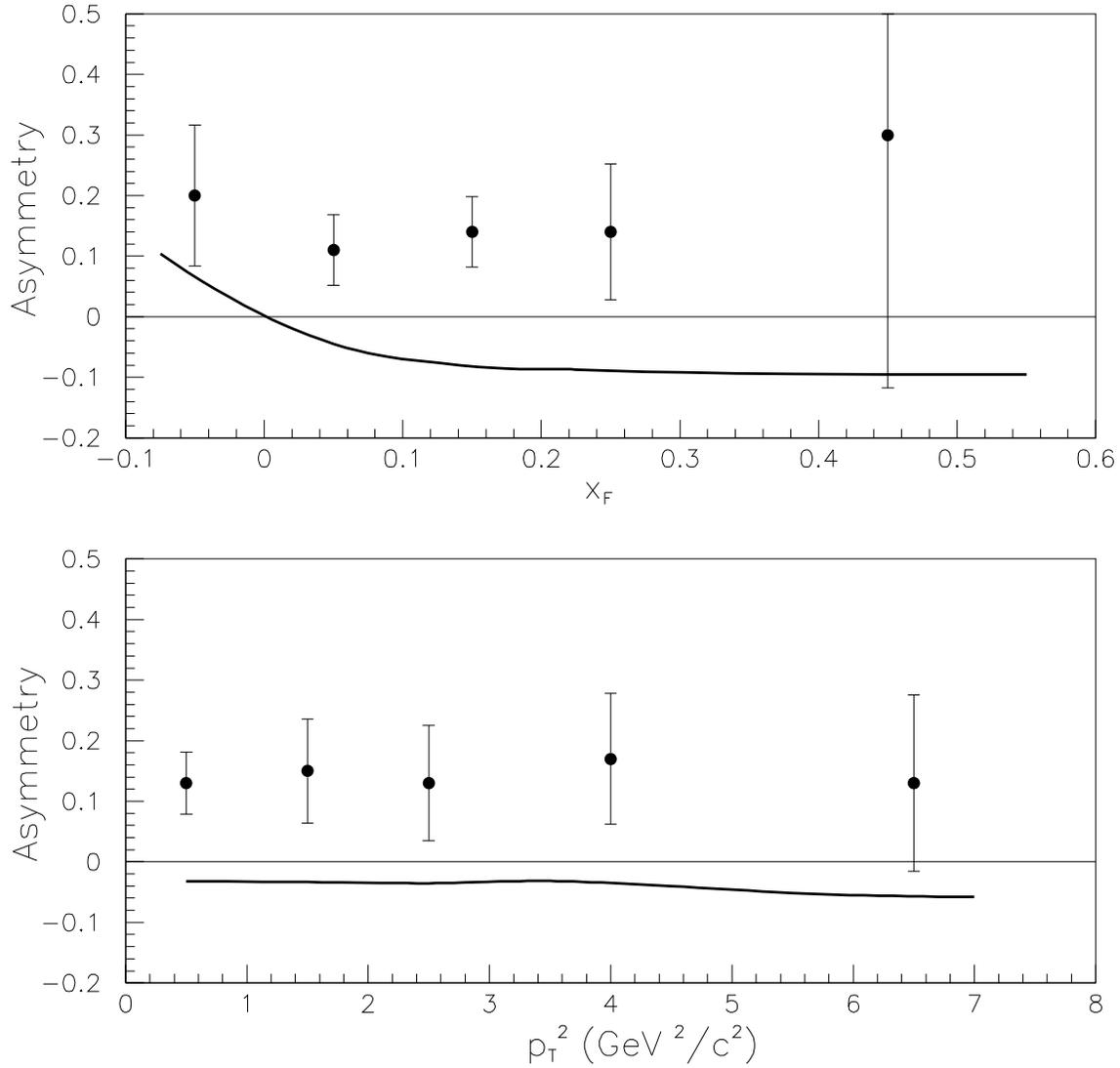}}
\caption{$\Lc - \Lcbar$ asymmetry as a function of 
$\xf$ (upper plot) and $\pt$ (lower plot). Full lines are the prediction of 
\pythi. The asymmetry for $x_F$ ($\pt$) is integrated 
over the $\pt$ ($x_F$) range of the data set. Error bars show 
the statistical and systematic errors added in quadrature.} 
\label{fig3}
\end{figure}

Acceptance-corrected asymmetries in the range $-0.1\leq \xf \leq0.6$ 
integrated over all $\pt\leq 8~({\rm GeV/c})^2 $, and 
in the $\pt\leq 8~({\rm GeV/c})^2$ range integrated from $-0.1$ to 
$0.6$ $\xf$, are shown in Fig. \ref{fig3} and listed in Table \ref{as}. 
The statistical errors given include those due to the number of observed
events and the number of MC events accepted. Typically, only for the
highest $\xf$ and $\pt$ bins is the error dominated by the MC uncertainty.
Also shown in Fig. \ref{fig3} are predictions from the default \pythia. 
The asymmetries predicted by \pyt are negative for $\xf>0$ and in all 
the $\pt$ range studied, while our data exhibit a positive asymmetry 
throughout the kinematical region studied. However, the data do not exclude a 
rise in the small negative $\xf$ region.


Sources of systematic uncertainties were checked. Among them we checked the 
effect of changing the parametrization of the signal and background shape, 
the effect of varying our principal selection criteria and the effect of 
a $2.5\; \%$ $K^-$ contamination in the beam.

The most significant effect comes from the parametrization of the background 
shape and the variation of our principal selection criteria. The effect of the $K^-$ 
contamination in the beam, which might produce a negative asymmetry in the 
$\xf>0$ region, was found to be negligible.

We also checked for contamination due to $D^{\pm}$ and $D_s^{\pm}$ 
decaying into 
$K\pi\pi$ and $KK\pi$ modes. We found that, by restricting our sample to have 
lifetimes between one and four $\Lcc$ lifetimes, we reduced any possible contamination to negligible levels.

The SELEX collaboration has measured an asymmetry in the $\xf>0.2$ region 
using different incident beam particles. Their
preliminary results for the $\pi^-$ beam indicate an asymmetry
of $A=(25\pm15)\%$ \cite{selex}.
The ACCMOR Collaboration \cite{accmor} 
has also measured the $\Lc-\Lcbar$ asymmetry over the $0\leq\xf\leq0.8$ 
region in 230 GeV/$c$ $\pi^-$--Cu interactions and found 
$A = (0.5 \pm 7.9)\%$, indicating no asymmetry, although with 
large uncertainty. Our measurement, when averaged over the 
$0\leq\xf\leq 0.6$ region, is $A=(12.3 \pm 3.7\pm1.6)\%$. 
No inconsistencies are observed in these 
results. The asymmetries measured by the three 
experiments and predictions of theoretical models are shown in Table~\ref{table2}.

In the forward $\xf$ region, leading particle effects 
are expected to cancel since, in this region, both particle 
and anti-particle share one valence quark 
with incident pions. The observed constant asymmetry could be due to 
different energy thresholds for particle and antiparticle in 
associated production of charmed mesons and baryons. A similar effect 
was observed in the hyperon production asymmetries measured in 
this experiment \cite{e791b}. Our data cannot exclude a rise in 
asymmetry in the $\xf<0$ region. 
The default \pythia model, which predicts a negative asymmetry 
over the range $\xf>0$, does 
not provide a good description of our data. The data show a trend 
similar to predictions of the two component models \cite{ic,rec}.

We have presented data on $\Lcc$ production 
asymmetries in both the forward ($\xf>0$) and backward ($\xf<0$) 
regions. The range of $\xf$ covered, $-0.1\leq \xf \leq 0.6$ allows 
the first simultaneous study of the $\Lcc$ production asymmetry 
in both the negative and positive $\xf$ regions. Our results show a 
uniform, positive asymmetry of $(12.7\pm3.4\pm1.3)\;\%$ 
after acceptance corrections over the entire kinematical range studied .

\begin{ack} 

We gratefully acknowledge the assistance of the staffs of 
Fermilab and of all the participating institutions. This research was 
supported by the Brazilian Conselho Nacional de Desenvolvimento Cient\'{\i}fico 
e Tecnol\'ogico, CONACyT (Mexico), the U.S.-Israel Binational Science Foundation, 
the U.S. Department of Energy and the U.S. National Science Foundation. 
Fermilab is operated by the Universities Research Associates, Inc., under  contract 
with the U.S. Department of Energy.

\end{ack}

\newpage

\newpage
\newpage
\begin{table}[t]
\begin{center}
\caption{ Measured $\Lc - \Lcbar$ asymmetry as a function of $\xf$ and $p_T^2$.}
\label{as}
\begin{tabular}{ccccccc} \hline \hline
$x_F$ bin & & Asymmetry & & $p_T^2$ bin & & Asymmetry \\ \hline
$-0.1$--$0.0$& &$0.20\pm0.10\pm0.06$& &$0.0$--$1.0$& &$0.13\pm0.05\pm0.01$ \\
$0.0$--$0.1$& &$0.11\pm0.05\pm0.03$& &$1.0$--$2.0$& &$0.15\pm0.07\pm0.05$ \\
$0.1$--$0.2$& &$0.14\pm0.05\pm0.03$& &$2.0$--$3.0$& &$0.13\pm0.09\pm0.03$ \\
$0.2$--$0.3$& &$0.14\pm0.10\pm0.05$& &$3.0$--$5.0$& &$0.17\pm0.10\pm0.04$ \\
$0.3$--$0.6$& &$0.30\pm0.39\pm0.15$& &$5.0$--$8.0$& &$0.13\pm0.14\pm0.04$ \\
\hline \hline
\end{tabular}
\end{center}
\end{table}
\begin{table}[t]
\begin{center}
\caption{Comparison of the $\Lc - \Lcbar$ asymmetries as measured by the 
E791, ACCMOR and SELEX Collaborations and predicted by three models. 
Experimental results in the $\xf>0$ region are for different $\xf$ ranges. 
See the text for details.}
\label{table2}
\begin{tabular}{ccccccc} \hline \hline
$\xf$ region  & E791 & ACCMOR~\cite{accmor} & SELEX~\cite{selex} & SF~\cite{lund} 
& IC~\cite{ic} & 2CR~\cite{rec} \\ \hline
$\xf<0$ & $0.20\pm0.10\pm0.06$   & $--$            & $--$         & $+$ & $+$ & $+$\\
$\xf>0$ & $0.123\pm0.037\pm0.016$ & $0.005\pm0.079$ & $0.25\pm0.15$ & $-$ & $0$ & $0$\\

\hline \hline
\end{tabular}
\end{center}
\end{table}
\newpage


\begin{thebibliography}{90}


\bibitem{mesons} E769 Collaboration (G.A. Alves, {\it et al.}),
Phys. Rev. Lett. {\bf 77}, 2388 (1996),  E769 Collaboration 
(G.A. Alves, {\it et al.}), Phys. Rev. Lett. {\bf 72}, 812 (1994), 
WA82 Collaboration  (M. Adamovich, {\it et al.}), 
Phys. Lett. {\bf B305}, 402 (1993), E791 Collaboration 
(E.M. Aitala, {\it et al.}), Phys. Lett.  {\bf B371}, 157 (1996), 
WA92 Collaboration (M. Adamovich, {\it et al.}),
Nucl. Phys. {\bf B495}, 3 (1997).

\bibitem{lund} B. Andersson, G. Gustafson, G. Ingelman and 
T. Sjostrand, Phys. Rep. {\bf 97}, 31 (1983).

\bibitem{ic}  R. Vogt and S.J. Brodsky, Nucl. Phys. {\bf B478}, 311
(1996).

\bibitem{rec} G. Herrera and J. Magnin, Eur. Phys. J. {\bf C2}, 477 (1998).

\bibitem{pyth} \pyt 5.7 and \jet 7.4 Physics Manual, CERN-TH-7112/93(1993).
 H. U. Bengtsson and T. Sjostrand, Computer Physics Commun.
{\bf 46}, 43 (1987) and T. Sjostrand, CERN-TH.7112/93 (1993).

\bibitem{e791} E791 Collaboration (E.M. Aitala \etal), Eur. Phys. J. {\bf C4}, 1 (1999); 
E791 Collaboration (E.M. Aitala, \etal), \PRL{76}, 364 (1996), and references therein; 
J.A. Appel, Ann. Rev. Nucl. Part. Sci. {\bf 42}, 367 (1992), and 
references therein; D.J. Summers \etal, Proceedings of the 
{\em XXVII${}^{\rm th}$ Rencontre de Moriond}, Electroweak Interactions
and Unified Theories, Les Arc, France (15-22 March, 1992) 417.

\bibitem{viteum} S. Amato, \etal, \NIM{A324}, 535 (1993).

\bibitem{e791-lamc-sigc} E791 Collaboration (E.M. Aitala, \etal), 
Phys. Lett. {\bf B379}, 292 (1996).

\bibitem{selex} James Russ, private communication; 
SELEX Collaboration (M. Iori \etal), Proceedings of the EPS-HEP99 
Conference, Tampere, Finland, July 1999, hep-ex/9910039.

\bibitem{accmor} ACCMOR Collaboration (S. Barlag, {\it et al.}), 
Phys. Lett. {\bf B247}, 113 (1990).

\bibitem{e791b} J.C. Anjos for the E791 Collaboration, Proceedings of the 
Hyperon'99 Conference, Fermilab (27-29 September, 1999), hep-ex/9912039.


\end{thebibliography}
\end{document}